\documentclass[12pt,preprint]{aastex}
\usepackage{color}
\usepackage{amsmath}
\usepackage[symbol]{footmisc}
\usepackage{graphicx}
\usepackage{dcolumn}
\usepackage{bm}
\usepackage{comment}
\usepackage{hyperref}
\usepackage{tikz}
\usepackage{changepage}
\usepackage{braket}

\begin{document}

\title{The Interiors of Uranus and Neptune:\\
 Current Understanding and Open Questions}

\author{
Ravit Helled$^{1}$ \& Jonathan J.~Fortney$^{2}$\\ 
{\small $^{1}$Center for Theoretical Astrophysics \& Cosmology, Institute for Computational Science, University of Zurich, Switzerland.\\
$^{2}$Department of Astronomy \& Astrophysics, University of California, Santa Cruz, CA, 95064 USA.\\
}
}

%
%
%
\begin{abstract}
Uranus and Neptune form a distinct class of planets in our solar system.  Given this fact, and ubiquity of similar-mass planets in other planetary systems, it is essential to understand their interior structure and composition.  However, there are more open questions regarding these planets than answers.  In this review we concentrate on the things we do not know about the interiors of Uranus and Neptune with a focus on why the planets may be different, rather than the same.  We next summarize the knowledge about the planets' internal structure and evolution.  
Finally, we identify the topics that should be investigated further on the theoretical front as well as required observations from space missions. 
\end{abstract}


%
%
%
%
%
%


\maketitle

\section{Introduction}
Uranus and Neptune are the outermost planets in the Solar System. 
These two planets raise great challenges to planetary scientists in terms of their formation history, evolution path, internal structure and composition, atmospheric dynamics, and many other areas.
Nevertheless, despite so many key questions around these planets, until recently, they have received relatively little attention. \par 

Modeling the internal structures of Uranus and Neptune is not simple. They represent a unique planetary class -- it is not possible to simply re-scale models of the terrestrial or gas giant planets, and many of the conclusions inferred on the planets' composition and internal structure surely reflect more the assumptions of the modeler than the reality. In a way, it is still unclear what the most reasonable assumptions should be when modeling these planets. 
Today, we know that even the gas giants, Jupiter and Saturn, have far more complex internal structures than had been assumed before the \emph{Juno} and \emph{Cassini} missions (Wahl et al. 2017, Fortney et al. 2019, Stevenson 2020, Mankovich 2020), and it certainly follows that if we now gained new detailed information about Uranus and Neptune, we would be similarly surprised.

The large number of observed exoplanets with masses and radii similar to those of Uranus and Neptune suggests that such planets are very common in the galaxy. 
Nevertheless, we still do not know as much about the ``ice giants'' as it is commonly assumed.  Before we state that a given exoplanet is similar to Uranus/Neptune, we first need to know what Neptune and Uranus are like. In addition, as we discuss below, it is still unclear how similar Uranus and Neptune are to each other, and we suggest that each planet should be investigated separately accounting for its unique features. 
\par

In the last several years efforts for designing space missions dedicated to the exploration of the ice giants have been made (e.g., Arridge et al.~2014, Masters et al.~2014, Mousis et al., 2018, Hofstadter et al.~2019, Fletcher et al.~2020). These efforts are very much ongoing, and we hope that a mission(s) to either or both of the ice giants will become a reality in the relatively near future.  

In this paper we focus on the interiors of Uranus and Neptune, and discuss the challenges they impose to the planetary science community. We list the key open questions and discuss the required developments in theory and observations.  
Other recent reviews and papers on the topic include Helled et al.~(2020), Guillot (2020), Helled \& Guillot (2018), Fortney \& Nettelmann, (2010), and references therein. 
This paper aims to be complementary to these recent publications, with a focus on open questions and what measurements would enable advances in our understanding.

\section{Not "Uranus and Neptune", but "Uranus" and "Neptune"}

It is clear that Uranus and Neptune represent a unique planetary class. They consist of H-He outer envelopes (and atmospheres) of about 10-20\% of their total mass, they are located in the outer part of the solar system, and they have similar masses, radii, and rotation periods. 
Therefore, as for Jupiter and Saturn, or Earth and Venus, but perhaps even more so here,
modelers have tended to investigate Uranus and Neptune together. 
However, each planet has clearly unique features, which justify a detailed investigation of each individual object. After all, just like the mentioned examples, there are clear differences between Jupiter and Saturn, as well as Venus and Earth, and by grouping them together, 
we may miss some of the key features that can reveal more information of the nature of each planet. In addition, always studying the two objects together may erode people's sense of the importance and uniqueness of each planet.
We therefore argue that each planet should be treated separately and uniquely.

\clearpage
Below we briefly discuss some of the key physical quantities of Uranus and Neptune, with a particular focuses on differences between the planets. 
Their fundamental physical properties are listed in Table 1. 
\begin{table}[h]
{\small 
\def\arraystretch{1.2}
\centering 
\begin{tabular}{lll}
\hline
\hline
{\bf Parameter} & {\bf Uranus} & {\bf Neptune}\\
\hline
Semi-major axis (AU) & 19.201 & 30.047 \\
Mass ($10^{24}$ kg) & 86.8127 $\pm$ 0.0040$^a$& 102.4126 $\pm$ 0.0048$^b$\\
Mean Radius$^*$ (km) & 25362 $\pm$ 7$^c$ & 24622 $\pm$ 19$^c$\\
Mean Density (g cm$^{-3}$) & 1.270 $\pm$ 0.001$^d$& 1.638 $\pm$ 0.004$^d$\\
R$_{\rm ref}$ (km) & 25,559$^a$& 25,225$^b$ \\
J$_2$ ($\times$10$^{6}$) &3510.68 $\pm$ 0.70$^a$& 3408.43 $\pm$ 4.50$^b$\\
J$_4$ ($\times$10$^{6}$) & -34.17 $\pm$ 1.30$^a$ & -33.40 $\pm$ 2.90$^b$\\
Rotation period$^*$ (Voyager) & 17.24 h$^e$& 16.11 h$^f$\\
1-bar Temperature (K) & 76 $\pm$ 2$^g$& 72 $\pm$ 2$^h$\\
Effective Temperature (K) & 59.1 $\pm$ 0.3$^i$ & 59.3 $\pm$ 0.8$^j$ \\ 
Intrinsic flux (J s$^{-1}$ m$^{-2}$) & 0.042 $\pm$ 0.045$^i$ & 0.433 $\pm$ 0.046$^j$\\
Bond Albedo $A$ & 0.30 $\pm$ 0.049$^h$ & 0.29 $\pm$ 0.067$^h$ \\
Axis tilt & 97.77  & 28.32  \\
\hline
\hline
\end{tabular}
\caption{Basic physical properties of Uranus \& Neptune. 
$^a$Jacobson, R.A. 2014. $^b$Jacobson, R.A. 2009.
$^c$Archinal et al. 2018. $^d$Calculated values and associated uncertainty derived from other referenced values and uncertainties in this table. The average density is computed using a  volume of a sphere with the listed mean radius. 
$^e$Desch et al., 1986 $^f$Warwick et al., 1989. $^g$Lindal et al. 1987. 
$^h$Lindal, 1992. 
$^i$Pearl et al. 1990 $^j$Pearl \& Conrath, 1991. 
$^*$Note that the rotation periods of the planets are not well determined as discussed in detail in Helled et al., 2010.  
R$_{\rm ref}$ is the reference equatorial radius in respect to the measured gravitational harmonics J$_2$ and J$_4$. } 
}
\label{tab:1}       
\end{table}

\subsection{Basic Physical Properties}
Uranus’ mass is slightly smaller than Neptune’s, but its radius is a bit larger. As a result, they differ in their mean densities which are 1.270 g cm$^{-3}$ and 1.638 g cm$^{-3}$ for Uranus and Neptune, respectively. This difference can already hint at different bulk compositions.  In addition, the inferred normalized moment of inertia (MoI) value of Uranus ($\sim$ 0.22) is smaller than that of Neptune ($\sim$ 0.24) , suggesting that Uranus is more centrally concentrated than Neptune (see Podolak \& Helled, 2012 and Nettelmann et al., 2013 for details).  

While the rotation periods of the planets are not well-determined, the \emph{Voyager} rotation periods suggest a difference in rotation period of 7\%.  \emph{Voyager 2} measurements of periodic variations in their radio signals and of fits to the magnetic fields of Uranus and Neptune imply rotation periods of 17.24 h and 16.11 h, respectively.  However, see Helled et al. (2010) for a discussion of why these values may be incorrect.  These differences in the most fundamental planetary properties already suggest that Uranus and Neptune should not be considered as "twin planets".

\subsection{Heliocentric Distance}
Uranus is located at $\sim$ 19 AU while Neptune is at $\sim$ 30 AU. While both of these locations are far from the sun, and represent the outer regions of the solar system, they have nearly 10 AU between them. For comparison, this is about the distance between Mercury and Saturn!  The factor of 2.5 difference in incident solar flux is larger than the difference between either Earth and Venus or Earth and Mars.

While at both of these radial distances it is clear that the two planets formed beyond the water and the CO$_2$ ice lines, 30 AU is closer to the nominal CO ice line (e.g., {\"O}berg et al., 2011). A difference of 10 AU could lead to substantial differences in the heavy element enrichments of the planets (see Mousis et al., 2020 and references therein). 
In addition, since the solid-surface densities at the two locations is different, the planetary growth history is is also expected to differ  (e.g., Helled \& Bodenheimer, 2014). 

It is possible that Uranus and Neptune formed much closer to the Sun, and even switched positions\footnote{With Neptune originally being closer to the Sun than Uranus, which could be supported by the fact that Neptune is slightly more massive than Uranus.} (e.g., Thommes et al. 1999; Tsiganis et al. 2005). This can partially resolve the formation timescale problem of Uranus and Neptune (e.g., Helled \& Bodenheimer, 2014).  However, a calculation that includes the planetary growth, accounting for the heavy-elements self-consistently, and that can lead to the predicted H-He-to-heavy elements ratios of either planet is still missing. 

\subsection{Axis tilt \& Satellite systems}
A distinct feature of Uranus is its axial tilt. 
This has typically been thought to be a result of a giant impact (Safronov 1966, Stevenson 1986, Podolak \& Helled, 2012, Kegerreis et al., 2018, 2019, Kurosaki \& Inutsuka, 2018, Reinhardt et al. 2020) although the tilt could also be a result of a spin-orbit resonance (Kubo-Oka \& Nakazawa 1995, Bou{\'e} \& Laskar 2010, Rogoszinski \& Hamilton 2020). 
Giant impacts might also be responsible for  some of the observed differences  of the planets such as their heat fluxes and internal structure\footnote{In fact, the small difference in masses could be a result of the impacts if indeed Neptune's impact was head-on, leading to the absorption of the entire impactor's mass, unlike an oblique impact on Uranus.} (Stevenson, 1986; Reinhardt et al., 2019).  In any case, no matter what the origin of the tilt is, it is clear that the seasons and temperature variations on the two planets are different, which affects the connection between the atmosphere and the deep interior and therefore the characterization of the planets (e.g., Guillot, 2020, Hueso  et al., 2020). 

Uranus has regular satellites, suggesting they formed from a circumplanetary disk.  It is yet to be determined whether the  circumplanetary disk was formed as a result of a giant impact (e.g., Kegerreis et al., 2018, Reinhardt et al., 2020, Ida et al., 2020) or a result of the gas accretion (e.g., Canup \& Ward, 2006). 
Unlike Uranus' regular satellites, the orbit of
Neptune's largest moon, Triton, is retrograde, implying that Triton was captured (e.g., Nesvorný et al., 2007), which perhaps destroyed any original regular satellite system.

While the differences in the axis tilt and satellite system could be a result of giant impacts with different conditions with the planets being more similar to each other shortly after formation, it would suggest different evolution histories. In addition, the origin of the moons of Uranus and Neptune is still being investigated, as well as the cause for the differences between the two planets. It is therefore clear that the unique features of each planet should be accounted for in their modeling.    

\subsection{Heat Flux and Albedo}

A strong indication for the dichotomy between  Uranus and Neptune comes from the far different energy balances of Uranus and Neptune.  Uranus's intrinsic power, as determined from the Voyager IRIS instrument is $42 \pm 47$ erg s$^{-1}$ cm$^{-2}$ (essentially a non-detection), while the value for Neptune is $433 \pm 46$, a value $\sim$5-10$\times$ larger (at $1\sigma$) (Pearl et al., 1990, Pearl \& Conrath 1991).  Uranus appears to be in equilibrium with solar insolation. 
Both planets have a similar Bond albedo, as determined by \emph{Voyager} data (Pearl \& Conrath 1991).  We discuss this in light of new data on Jupiter's Bond albedo in a later section.

\subsection{Atmosphere: Activity, Depth of Winds, Composition}
Images of Uranus and Neptune imply that Uranus' atmosphere has only very few features while Neptune's atmosphere seems more active and complex including storms and vortexes. This could either be a result of different internal heat fluxes, the different orbital properties, as well as the different seasons due to the different axial tilt. It is desirable to understand the origin of the different atmospheric activity. 

In terms of atmospheric winds, the two planets seem to have a similar wind profile characterized by a strong eastward jet at the equator. The wind velocities are referenced in comparison to the underlying assumed rotation periods which are likely more uncertainly than typically appreciated. It was shown in Helled et al.~(2010) that the inferred flattening  of the planets are inconsistent with the \emph{Voyager} radio periods, and new rotation periods which are more consistent with the data have been suggested. With these modified periods (of  16.58 h for Uranus and 17.46 h for Neptune) the wind velocities on the two planets are much more similar and are slower than previously thought. As a result, the wind speeds are not well known but are expected to be of the order of a couple 100 m/s. The penetration depths of these winds are unknown, but are thought to be as deep as 1100 km for both planets, as estimated from their gravity data (Hubbard et al. 1991, Kaspi et al., 2013). 

The atmospheres of both planets are so cold that most volatile species have condensed into clouds far below the visible atmosphere.  In addition, the noble gases would only be detectable via an entry probe with a mass spectrometer.  The atmospheres of Uranus and Neptune are mostly H-He (by particle numbers) with smaller fractions of heavier elements. The only heavy element with a well-determined composition is carbon, in the form of methane, CH$_4$.  Even this measurement is problematic, as CH$_4$ partially condenses at such cool temperatures.  Recent assessments of the CH$_4$ abundances are Karkoschka \& Tomasko (2011) for Uranus and Sromovsky et al. (2011) for Neptune.  See Atreya et al. (2020) for a detailed review and discussion of how latitude-dependent condensation effects these measurements. 

Compared to the solar carbon abundance, these carbon values are $85\pm 15$ for Uranus and $89 \pm 22$ for Neptune.  For comparison, the methane abundance in Saturn from \emph{Cassini} spectroscopy is $9.9 \pm 0.4$ (Fletcher et al., 2009), and for Jupiter it is $4.4 \pm 1.1$ from the \emph{Galileo} entry probe (Wong et al., 2004).  This may suggest that there is a strong anti-correlation between atmospheric metallicity and giant planet mass.  However, the expected  very different formation locations of the four planets make this direct interpretation difficult.

The previously mentioned condensation of CH$_4$, and other volatiles like H$_2$S and H$_2$O, is important for reasons beyond understanding composition.  Condensation into clouds removes these molecules from the gas phase, which alters the mean molecular weight of the atmosphere across the relatively narrow thickness of the cloud.  This may lead to a number of regions of the middle and deep atmosphere becoming superadiabatic, as a steeper temperature gradient would be needed to drive convective motion.  The connection with the planetary interior is that these atmospheric superadiabatic regions would lead the deep interior to be hotter than previous simple estimates.  Much additional discussion can be found in Guillot (1995), Guillot (2020), Fletcher et al., (2020), Leconte et al., (2017), and Friedson \& Gonzalez (2017).



\section{Interior Models}
The basic idea of planetary modeling is as follows: given the measured physical properties of a planet  (mass, radius, rotation rate, gravity field, etc.) a structure model is developed to reproduce the observed properties. The density profiles that fit the data can teach us about the planetary composition and its depth dependence. 
The more measurements we have, and the more accurate they are, the better the internal structure is determined. 
However, accurate measurements are insufficient to \emph{uniquely infer} the planetary structure and composition, given the degenerate nature of the problem. Even if we precisely knew all the fundamental properties of a planet, such as its mass, shape, gravitational and magnetic fields, there would still be ambiguity in determining the composition and internal structure. This is because there is more than one solution for the planetary density profile that can satisfy all the observational constraints. 
In addition, structure models suffer from "theoretical uncertainties" that are linked to the  EOS, composition, energy transport mechanism, and structure assumed by the modeler.
\par

Nevertheless, the available data can be used to exclude certain solutions, and identify solutions that seem to be consistent with complementary knowledge about the planets such as their formation location (far from the sun, beyond certain ice lines) and predicted composition, magnetic fields (e.g., the need for convection and high-enough electrical conductivity), and the behaviour of elements at high pressures/temperatures that can guide structure models (see Helled et al., 2020 and references therein for details). 
However, at the moment, some key observed properties of Uranus and Neptune are not well determined. 
This is in particular notable now when the two planets are compared to Jupiter and Saturn which were explored by various spacecraft through the years including the recent visits by \emph{Juno} and \emph{Cassini} which provided unprecedented  accurate measurements of their gravity fields. 
For Uranus and Neptune the gravitational moments are determined only up to fourth degree ($J_2, J_4$), with a relatively large uncertainty, and detailed information on their atmospheric composition, planetary shape, and rotation periods is missing (e.g., Helled et al. 2010).  It should be noted that a recent study provided a new estimate for Neptune's second gravitational moment, $J_2 =  3409.1 \pm 2.9 \times 10^{-6}$ (Brozović et al., 2020). This value is consistent with the value of $3408.43 \pm 4.5 \times 10^{-6}$  inferred by Jacobson (2009).

 \begin{table}[h!]
{\small 
\def\arraystretch{1.3}
\centering 
\begin{tabular}{lllll}
\hline
\hline
& {\bf Jupiter} & {\bf Saturn} &{\bf Uranus} & {\bf Neptune}\\
\hline

J$_2$ old ($\times$10$^{6}$) & 14696.43 $\pm$ 0.21 & 
16290.71 $\pm$ 0.27  & 3516 $\pm$  3.2 & 3539 $\pm$ 10  \\

J$_2$ new ($\times$10$^{6}$) & 14696.572 $\pm$ 0.014 & 16290.557 $\pm$ 0.028  & 3510.68 $\pm$ 0.70  & 3408.43 $\pm$ 4.50  \\
J$_4$ old ($\times$10$^{6}$) & -587.14 $\pm$ 1.68  & -935.8 $\pm$ 2.8  & -35.4 $\pm$ 4.1  & -28 $\pm$ 22 \\
J$_4$ new ($\times$10$^{6}$) & -586.609 $\pm$ 0.004 & -935.318 $\pm$0.044 & -30.44 $\pm$ 1.02 & -33.40 $\pm$ 2.90\\
\hline
\hline
\end{tabular}
\caption{Gravity data from Figure 1. The old and new gravity data for Jupiter are taken from Jacobson (2003) and Iess et al.~(2018), respectively. For Saturn old gravity data corresponds to Jacobson (2006) and the updated one from Iess et al.~(2019). The gravity data for Uranus are from French et al.~(1988, "old") and Jacobson (2014, "new"), and for Neptune from Tyler et al.~(1989,"old") and Jacobson (2009, "new"), respectively. Note that "old" data correspond to the latest data before the most recent, i.e., "new" data.} 
}
\label{tab:1}       
\end{table}

Figure 1 shows the gravity measurement ($J_2$, $J_4$) uncertainties of the outer planets. Shown are the improvements of new data in comparison to older ones. The values of the  gravitational moments used for this figure are listed in Table 2. Thanks to the \emph{Juno} and  \emph{Cassini} Grand Finale measurements, the uncertainties in the gravity data decreased dramatically for Jupiter and Saturn. Such accurate data push structure models to the next-level of complexity (see Helled, 2018 for review). Unfortunately, it is clear that the quality of the data for Uranus and Neptune is orders of magnitude worse. As a result, the compositions and internal structures of these planets are poorly determined. 
It is clear that measuring the gravity fields of Uranus and Neptune accurately is desirable; such accurate measurements can only be done via a Juno-like spacecraft  which orbits the plant several times, with the orbits being polar, covering different regions, and reach close to the planet. We therefore strongly support orbiter missions to the ice giants (e.g., Fletcher et al., 2020 and references therein). 

Uranus and Neptune are special from a theoretical modeling perspective because they are clearly different from the terrestrial planets and the gas giants. Therefore, it is not clear what the best approach to take is when modeling their interiors. 
Contrary to most published models, and corresponding artists conceptions, it is unlikely that they are fully differentiated objects with distinct layers.  Although it is often assumed that they are water-rich, their composition is poorly constrained as discussed in Helled et al.~(2011, 2020). Also, since they are not H-He dominated, the chosen materials and their distribution significantly affects the inferred composition. Planets in this mass regime are in fact most sensitive to the assumed planetary internal structure and the equation of state (EOS) used by the modeler (e.g., Baraffe et al., 2008, Vazan et al., 2013).  \\

Planet formation models suggest that deep interiors of gaseous planets are expected to consist of composition gradients, although this is just started to be included in current structure models.
During the planetary growth phase the accreted heavy elements (pebbles and/or planetesimals) are dissolved in the envelope already when the core mass reaches a few Earth masses (M$_{\oplus}$). An example for such a gradient has been presented in Helled \& Stevenson (2017) who also note that the more gradual structure is associated with a lower solid surface density.  As a result, the primordial interior of Uranus and Neptune are expected to have a large region with composition gradients. Figure 2 shows the heavy-element mass fraction as a function of planetary mass during the planetary growth when assuming two different solid surface densities. 
 These curves are not meant to represent proto-Uranus/Neptune but to demonstrate the sensitivity of the inferred primordial structure to the assumed formation environment. As a result, a better understanding of the evolution and internal structure of Uranus and Neptune, could help us to better understand their origin (see  Helled et al.~2020 for further discussions). 
The heavy-element accretion rate is affected by the (assumed) solid surface density $\sigma$, and this in return affects the subsequent growth and the gas accretion rate. As a result, not only do the planets form in different timescales but they also have very different final compositions. It is clear that for the case with lower $\sigma$ the planet is more H-He rich. In both cases, the planets are expected to have primordial internal structures with composition gradients  (see Helled \& Stevenson, 2017 and Helled et al.~2020 for further discussions). 
 Such primordial composition gradients are expected to evolve on a timescale of 10$^9$ and convective mixing could lead to an outer region that is convective. We suggest that future studies should investigate different formation paths that can lead to planets that are similar to Uranus and Neptune.   
 
Since both Uranus and Neptune are located in the outer parts of the Solar System where the solid-surface density is low, even if they migrated outward they are likely to have formed further than Jupiter and Saturn, and therefore, have composition gradients. Composition gradients are also predicted for the gas giants.  There is compelling seismic evidence for a deep composition gradient in Saturn (Fuller 2014) and interesting (but more indirect, since it is from the gravity field) for Jupiter (Wahl et al., 2017, Vazan et al., 2018, Debras \& Chabrier, 2019).

Interior models of Uranus and Neptune assuming adiabatic temperature profiles with distinct layers have been presented by Nettelmann et al.~(2013). In these models the planets are assumed to consist of a rocky core surrounded by a water layer, and a H-He atmosphere with a given metallicity. These models use physical EOSs to model different materials. While these models are in some way "more physical" they are rather sensitive to the model assumptions.  Perhaps most troubling is that such distinct-layer models tend to predict extremely high water-to-rock ratios for both planets. The inferred Uranus and Neptune models of Nettelann et al.~(2013) predict that the water-to-rock ratio is 19-35 times for Uranus, and 4-15 times for Neptune, with the total H-He mass is typically 2 and 3 M$_{\oplus}$ for Uranus and Neptune, respectively. The exact estimates are highly model-dependent, and are sensitive to the assumed composition, thermal structure (which depends on the assumed heat transport mechanism), and rotation rate. 
It is important to note that such structure models with distinct layers of different compositions are likely unrealistic since rock and water can be mixed as well as water and hydrogen (Soubiran \& Militzer, 2015, Soubiran et al. 2017). In addition, the extremely high water-to-rock ratios are not observed in any solar-system object, and as discussed above there is a strong indication from formation and evolution models that the planets consist of composition gradients.  Indeed, models where rock is more gradually distributed within the planet go in the direction of alleviating the high water-to-rock ratio problem. As a result, such 3-layer models are unlikely to properly represent the interiors Uranus and Neptune. 

Another approach for interior modeling is to take a more unbiased view on the internal structures of the planets.  This is by producing empirical density profiles (e.g., Marley et al., 1995, Podolak et al., 2000, Helled et al., 2011). 
In that case the planetary density profile is represented by a series of random steps in density or a mathematical function, and all the density profiles that match the observed properties are inferred. These can then be interpreted using physical EOS. 
While the inferred density profiles are not based on knowledge of the behavior of elements at high pressures and temperatures, and can therefore might be viewed as "less physical", they can probe solutions that are missed by the standard models, in particular, solutions that represent more complex interiors (e.g., composition gradients) with various temperature profiles (e.g., sub- and super- adiabatic). 
\par 

For example, empirical models of Uranus and Neptune using 6th order polynomials to represent the density profile suggest that both planets can have a continuous density profile in which there is a gradual increase of the heavier material toward the center (Helled et al.,~2011), and that the  overall metals mass fraction of the planets is 0.75-0.92 and 0.76-0.9 for Uranus and Neptune, respectively.
It was also shown that the planetary interiors are not necessarily water-rich, and that the measured gravitational field can be reproduced also if the planets are assumed to be rock-dominated. Finally, it was shown that the inferred densities are consistent with compositions gradients, and therefore with non-adiabatic temperature profiles (see Helled et al., 2020 for discussion).  
Representative density profiles of Uranus (left) and Neptune (right) inferred by various studies are presented in Figure 3. 
\par

Recently, it was shown that Uranus' low luminosity can be a result of a combination of primordial composition gradients that inhibit convection and a low planetary luminosity post-formation (Vazan \& Helled, 2020). 
In this scenario the deep interior of Uranus can be very hot, which also suggests that the planet could consist of more refractory materials, since with higher internal temperatures one is able to include more rocks within a given model (since the high temperature makes the rocks lower density). 
In addition, a stable (to convection) composition gradient implies that Uranus' current-state internal structure has not significantly evolved, and could be used to guide planet formation models. 
Although these evolution models are not designed to fit the gravity data exactly, they can reproduce the basic measured planetary properties. 

Given that a wide range of structures are possible to fit the gravity fields of each planet, this naturally begs the question regarding how large the possible range of solutions really is?  Movshovitz et al. (2020a) presented a Bayesian MCMC-driven approach to exploring the full range of interior models for a giant planet, given the gravity field.  The first application was to Saturn, but in Figure \ref{fig:naor} we show preliminary work from Moshovitz et al.~(2020b) that shows all the density profiles that fit the current gravity fields of Uranus and Neptune.  These models use an 8th order polynomial for most of the planet, which allows for relatively steep changes in density vs.~radius, if such changes are needed to fit the gravity field constraints.  As was previously demonstrated (Marley et al. 1995, Helled et al. 2011), for the interiors of Uranus and Neptune, there is no need not be distinct layers. 

Currently, the gravity field of Uranus is better constrained than Neptune (see Table 1).  This manifests itself as wider range of allowed interior structures in Neptune in Figure \ref{fig:naor}.  This can be seen most clearly from 0.3 to 1.0 planetary radii (see in particular the insets) where Uranus is better-constrained. This is because 
the behavior of the contribution functions of the gravitational moments, which describe how the different layers in the planetary interiors contribute to the  gravitational moments. For both Uranus and Neptune the innermost 30\% of the planets is not well sampled by the gravitational harmonics. As a result, conclusions about the innermost part of the planets must be inferred indirectly (see Helled et al., 2011 for details). 
In addition, the error bar on $J_6$ on either planet is so large that it provides no real additional constraint. 

As the atmospheric composition of Uranus and Neptune can be viewed as windows to their deep interiors, improved understanding of their atmospheres can further constrain their formation path,  thermal evolution, and internal structures (see Helled et al., 2020 for details).     
For instance, the pollution of the protoplanetary atmospheres with heavy elements such as water, ammonia, and methane can significantly affect the cooling of the growing planet and therefore its formation and evolution paths (Kurosaki \& Ikoma, 2017). In addition, the elemental abundances can reveal information on the formation locations of the planets and/or the composition of their building blocks.  
Therefore, it is clear that measuring the atmospheric composition of Uranus and Neptune is also desirable.

\section{Energy Balance and Evolution}
Models that aim to understand the amount of thermal flux coming from the planetary interior today are an important complement to models of a planet's current density structure.  In principle, one should aim for a coherent picture of interior structure and thermal evolution over time, tied to the planet's formation.

Significant thought and modeling work have gone into trying to explain the dramatic difference in the heat flux between the two planets.  The ``standard story''  (e.g., Fortney et al. 2011) has been that Neptune's flux value is basically what one would expect for the cooling of a 3-layer model with an adiabatic interior, while the value for Uranus is far too low.  Investigations have shown that either assumed barriers to convection (Nettelmann et al. 2016) or composition gradients suggested by formation models (Vazan \& Helled, 2020) can radically alter Uranus's cooling history, leading to a low intrinsic flux today, along with a hot interior that his unable to efficiently cool off.

At the same time, it has been seen that the standard story may not be the correct one.  Recently, due to \emph{Cassini} Mission Jupiter fly-by data, Jupiter's Bond albedo and intrinsic flux determination were significantly updated (Li et al., 2018) -- the Bond albedo increased from 0.343 to 0.503 (a 46\% increase), and the intrinsic flux determination increasing by 38\%, based on a combination of the new Bond albedo data and improved measurements of the planet's total thermal emission.  This at least brings about the possibility that \emph{Voyager}-derived energy balance values for the other 3 giant planets (Pearl \& Conrath, 1991) could be in need of substantial revision. Figure \ref{fig:alb} shows the Bond albedo values for the planets.  We note that the Bond albedo can only be determined by a space mission that observes scattered sunlight over a wide range of wavelengths and phase angles.

Such energy balance revisions can radically alter the standard Uranus/Neptune picture.  Furthermore, the accuracy of input physics is always improving, and it is essential to revisit models as physics improves.  Scheibe et al. (2019), assuming the \emph{Voyager} Bond albedos for each planet and a time-evolving solar luminosity, find that Neptune is actually \emph{overluminous} compared to the expectation of an adiabatic cooling model, and Uranus is still underluminous, but not be as much as previously thought.  A modestly higher Bond albedo for Uranus (from 0.3 up to 0.4) would make Uranus's model adiabatic cooling history fit with observations.  Neptune would appear to be overluminous, with \emph{any} Bond albedo.  This is a "flip" from previous work, marking Neptune, rather than Uranus, as the "odd" planet.

Given this wide variety of evidence, it is now becoming ever clearer that Uranus and Neptune should not be modeled assuming adiabatic interiors with distinct layers, and the field is certainly moving in that direction (Nettelmann et al. 2016, Podolak et al., 2019, Scheibe et al.~2019, Vazan \& Helled, 2020).  An important path forward is expected to come from detailed models of planet formation, and then evolving such models over 4.5 Gyr of time, to assess their current structure and heat flux today, compared to observations.

\section{Should we keep calling Uranus and Neptune the "ice giants"?}
Often Uranus and Neptune are referred to as the "ice giants." The origin of the name is probably linked to the mean densities of the planets which are comparable to the density of water, and due to the fact that they are located at large radial distances where volatile materials can condense to form ices (water, ammonia, methane). 

However, we actually do not know if the compositions of Uranus and Neptune are dominated by these materials, and even Pluto seems to consist of more rock than ice (McKinnon et al., 2017). Indeed it was shown that the observed properties of the planets can be fit also with a rock-dominated composition (Helled et al., 2011), and recently, it has been suggested that Neptune could be a "rock-giant" based on measured atmospheric abundances (see Teanby et al., 2020 and references therein for further details). 
Also, although the argument that the planets must consist of large fractions of water to have  high enough electrical conductivities (ionic/super-ionic water) to generate their magnetic fields is convincing (e.g., Redmer et al. 2011), it is yet to be determined how much water is required and whether other materials could contribute to the ionic interior. 

In addition, even if the planets have substantial amounts of volatile materials (e.g., water), in their deep interiors, the physical state of the material would not be in a solid state, and therefore it is inappropriate to describe the materials as "ices" since they would be in the liquid (fluid) state. This is in fact also true for Jupiter and Saturn which are called the "gas giants", because their composition is dominated by hydrogen (H), although the material in their deep interiors is not in the gaseous phase. 
Similarly, we suggest that calling Uranus and Neptune ``ice giants" is rather misleading. This name biases the community to think of these planets as being water- (volatiles) dominated and also gives the wrong impression for the physical state of the material in their deep interiors. 
We therefore propose that naming Uranus and Neptune "sub-giants" or "outer-giants" instead of "ice giants" is more appropriate\footnote{We note that the earliest recorded use of "ice-giant" we can find is in the introduction to a 1978 NASA report about the Mariner 10 mission to Mercury (Dunne \& Burgess, 1978).}.

\section{Summary \& Future Plans}
Uranus and Neptune remain mysterious planets and it is clear that further exploration of these planets theoretically and observationally is needed. 
Key fundamental questions regarding Uranus and Neptune remain open, such as:
Key questions regarding Uranus and Neptune include:
\begin{itemize}
\item[+]  How do planets like Uranus and Neptune form?
\item[+]   How do these planets evolve?
\item[+]   What are the compositions and internal structures of Uranus and Neptune?
\item[+]  Are Uranus and Neptune volatile-rich?  
\item[+]   What parts of each planet are superadiabatic?
\item[+]  How different are Uranus and Neptune? What is the origin of these differences? 
\item[+]   How is the magnetic field generated? 
\end{itemize}

The collection of these important open questions ensures that investigating Uranus and Neptune in the near and far future will be extremely rewarding.  
Although the development of new theoretical models is crucial, it is clear that significant progress in our knowledge cannot be achieved without more data.  
We argue that missions to Uranus and Neptune are essential. In particular we emphasize the importance of accurate measurements of the planets' {\bf gravitational fields} (preferable with several \emph{Juno}-like polar orbits). 
Better gravity data will allow us to exclude certain solutions for the interiors, and can assist us in understanding the differences between Uranus and Neptune. 
We will then be able to reduce the parameter space of possible internal structures.
Better determining the structure and potential variability of the planetary {\bf magnetic fields} will allow us to further constrain the planetary structure via the required conditions to sustain a dynamo (electrically conducting material + convection),  as discussed in detail in the complementary chapter by Soderlund \& Stanley (2020) in this issue. 
At the same time, measuring the 
{\bf atmospheric composition} from an entry probe can be used to better understand the connection between the atmosphere and the deep interior (e.g., Guillot, 2020) as well as the  origin of the planets (e.g., Mousis et al., 2020).
In addition, determining the {\bf rotation rates} of the planets helps to tighten constraints on interior models.  Better establishing each planet's {\bf Bond albedo} and {\bf thermal fluxes} can be used to further constrain structure and evolution models, respectively. 
Due to the complex nature of planets, it is clear that having one measurement, even if very accurate, is insufficient to break the degeneracy of structure models and reveal the true nature of the planets. 
What is needed is a comprehensive investigation of each of the two planets using different measurements so we can slowly put the different pieces of the puzzle together until we better understand Uranus and Neptune.    


In the nearer future, before potential space missions, some progress is envision in the following fonts. 
Further improvements in EOS calculations and experiments of volatile materials such as water, ammonia, and methane, their mixtures, as well as their mixtures with rock or with hydrogen (and helium), are essential.  This could build on recent advances (e.g., Bethkenhagen et al., 2017, Millot et al., 2019). 
Understanding the behaviour of the materials, and their mixtures, we think exist in Uranus and Neptune at high pressures and temperatures, will allow us to exclude certain compositions, and can guide us in terms of the model assumptions: i.e., what materials are likely to be mixed vs. differentiated and what are the important chemical interactions that should be considered.

A longer-term investment, that could start now, would be further investigations towards gains that could be made by seismology of Uranus and Neptune.  Building on important work for Jupiter and Saturn over the past decade (Gaulme et al., 2011, Fuller 2014, Mankovich et al., 2019), further theoretical and observation-driven work could be imagined.  This includes assessing the potential of ``ring seismology of Uranus" using detailed observations of its ring system (Hedman et al., 2014), or also long time baseline photometry (Rowe et al., 2017).

Detailed and accurate measurements of atmospheric abundances of giant and intermediate-mass exoplanets will also be valuable. An overview of the variation in atmospheric composition of gaseous exoplanets in the mass/size range of Uranus and Neptune, and its connection to the host star's properties, in concert with determinations of the planetary mean density will allow us to understand the nature of gaseous-rich intermediate mass planets.

\par
%
%
%
%
%
%
%

\section*{Acknowledgments}
We thank the two anonymous referees for valuable comments.  RH thanks D.~Soyuer and B.~Neuenschwander for technical support. RH also acknowledges support from SNSF grant 200020\_188460. JJF acknowledges support from NASA grant NNX16AI43G and NSF-AAG grant 1908615. 

\section*{References}
\small{
Arridge, C. S., Achilleos, N., Agarwal, J., (2014), The science case for an orbital mission to Uranus: Exploring the origins and evolution of ice giant planets {\it Planetary \& Space Science}, 104, 122.\\
\indent Baraffe, I., Chabrier, G., \& Barman, T.~(2008). Structure and evolution of super-Earth to super-Jupiter exoplanets: I. heavy element enrichment in the interior. {\it A\&A}, 482, 315. 

\indent Bethkenhagen, M., Meyer, E. R., Hamel, S., Nettelmann, N., French, M., Scheibe, L., Ticknor, C., Collins, L. A., Kress, J. D., Fortney, J. J., \& Redmer, R., (2017), Planetary Ices and the Linear Mixing Approximation, {\it ApJ}, 848, 67.  

\indent Boué, G.~\& Laskar, J., (2010). A Collisionless Scenario for Uranus Tilting. {\it ApJ}, 712, L44.  

\indent
Brozović, M.,  Showalter, M.~R.,  Jacobson, R.~A., French, R.~S.,   Lissauer, J.~J.~\&  de Pater, I., (2020). 
Orbits and resonances of the regular moons of Neptune. 
{\it Icarus}, 338, article id. 113462.  
\indent
Debras, F.~\& Chabrier, G., (2019). New Models of Jupiter in the Context of Juno and Galileo. {\it ApJ}, 872, 100.  \indent

%
Desch, M.~D., Connerney, J.~E.~P.~\& Kaiser, M.~L.,  (1986). The rotation period of Uranus. {\it Nature}, 322, 42.  \indent
 
Dunne, J. A.~\& Burgess, E., (1978). The voyage of Mariner 10: mission to Venus and Mercury. NASA Special Publication, 424.  \indent

 Fletcher, L, Helled, R., Roussos, E.~et al., (2020). Ice Giant Circulation Patterns: Implications for Atmospheric Probes. {\it Space Science Reviews}, 216, 21.  \indent
 
 Fletcher, L. N., Orton, G. S., Teanby, N. A., Irwin, P. G. J., \& Bjoraker, G. L., (2009). Methane and its isotopologues on Saturn from Cassini/CIRS observations. {\it Icarus}, 199, 351.  \indent
 
Fortney, J.J., Ikoma, M., Nettelmann, N., Guillot, T.~\& Marley, M.S.
\newblock {Self-consistent Model Atmospheres and the Cooling of the Solar
  System Giant Planets}.
\newblock {\em ApJ}, 729, 32, (2011). \indent
 
Fortney, J.J.~\& Nettelmann, N., (2010). 
\newblock {The Interior Structure, Composition, and Evolution of Giant
  Planets}.
\newblock {\em Springer Space Science Reviews}, 152, 423.   \indent
 
Fortney, J. J., Helled, R., Nettelmann, N., Stevenson, D. J., Marley, M. S., Hubbard, W. B. \& Iess, L. The Interior of Saturn, In Kevin H. Baines, F. Michael Flasar, Norbert Krupp, Tom Stallard, editors, {\em Saturn in the 21st Century}, pages 44--68. Cambridge University Press, (2019).  \indent

French, R. G.,  Elliot, J. L.,  French, L. M.~et al., (1988).  Uranian ring orbits from earth-based and Voyager occultation observations.  
{\it Icarus}, 73, 349.  \indent

Friedson, A. J.~\& Gonzales, E. J., (2017), Inhibition of ordinary and diffusive convection in the water condensation zone of the ice giants and implications for their thermal evolution, {\it Icarus}, 297, 160. \indent
 
Fuller, J. (2014). Saturn ring seismology: Evidence for stable stratification in the deep interior of Saturn. {\it Icarus}, 242, 283.  \indent
 
Gaulme, P., Schmider, F.-X., Gay, J., Guillot, T.~\& Jacob, C.,  (2011). Detection of Jovian seismic waves: a new probe of its interior structure. {\it A\&A}, 531, A104.  \indent
 
Guillot, T., (1995). Condensation of Methane, Ammonia, and Water and the Inhibition of Convection in Giant Planets, {\it Science}, 269, 1697. \indent

%
Guillot, T., (2020). Uranus and Neptune are key to understand planets with hydrogen atmospheres. ESA Voyage 2050 White Paper, arXiv:1908.02092.   \indent
 
Hanel, R., Conrath, B., Herath, L., Kunde, V., Pirraglia, J., (1981). Albedo, internal heat, and energy balance of Jupiter: preliminary results of the voyager infrared investigation. {\it Journal of Geophysical Research}, 86, 8705. \indent
 
Hanel, R. A., Conrath, B. J., Kunde, V. G., Pearl, J. C., Pirraglia, J. A., (1983). Albedo, internal heat flux, and energy balance of Saturn. {\it Icarus}, 53, 262. \indent

Hedman, M. M., French, R. G.~\& McGhee-French, C. (2014). What can Uranus' Rings Tell Us About Uranus' Internal Structure?.  Workshop on the Study of the Ice Giant Planets, 1798, 2013.  \indent
 
Helled, R., Anderson,  J.~D.~\& Schubert G.~(2010). Uranus and Neptune: Shape and rotation. {\it Icarus}, 210, 446.  \indent
 
Helled, R., Anderson, J.~D., Podolak, M.~\& Schubert G.~(2011). Interior models of Uranus and Neptune. {\it ApJ}, 726,15.  \indent
 
Helled, R., Bodenheimer, P., (2014). The Formation of Uranus and Neptune: Challenges and Implications for Intermediate-mass Exoplanets.  {\it ApJ}, 789, 69.  \indent

Helled, R.~\& Stevenson, D.J.~, (2017). The Fuzziness of Giant Planets' Cores. {\it ApJ}, 840, L4.  \indent
 
Helled, R. \& Guillot, T., Internal Structure of Giant and Icy Planets: Importance of Heavy Elements and Mixing. Handbook of Exoplanets, ISBN 978-3-319-55332-0. Springer International Publishing AG, part of Springer Nature, 2018, id.44. (2018).  \indent
 
Helled, R., Nettelmann, N.~\& Guillot, T., (2020). Uranus and Neptune: Origin, Evolution and Internal Structure. {\it Space Sci Rev}, 216, 38.  \indent
 
Hofstadter, M., Simon, A., Atreya, S., Banfield, D., Fortney, J. J., Hayes, A., Hedman, M., Hospodarsky, G., Mandt, K., Masters, A., Showalter, M., Soderlund, K. M., Turrini, D., Turtle, E., Reh, K., Elliott, J., Arora, N., Petropoulos, A.. Uranus and Neptune missions: A study in advance of the next Planetary Science Decadal Survey. {\it Planetary \& Space Science}, 177, 104680. \indent
%
%
%

Hubbard, W. B., Nellis, W. J., Mitchell, A. C., Holmes, N. C., Limaye, S. S., \& McCandless, P. C., (1991). Interior Structure of Neptune: Comparison with Uranus. {\it Science}, 253, 648. \indent

Hueso, R., Guillot, T.~\& Sánchez-Lavega, A., (2020). 
Atmospheric dynamics and convective regimes in the non-homogeneous weather layers of the ice giants. {\it RSTA}, submitted.  \indent
 
Ida, S., Ueta, S., Sasaki, T.~et al. (2020). Uranian satellite formation by evolution of a water vapour disk generated by a giant impact. {\it Nat Astron}.  https://doi.org/10.1038/s41550-020-1049-8 \indent
 
Iess, L., Folkner, W. M., Durante, D.~et al., (2018). Measurement of Jupiter's asymmetric gravity field. {\it Nature}, 555, 220.  \indent
 
Iess, L.,  Militzer, B.,  Kaspi, Y.~et al., (2019).  Measurement and implications of Saturn's gravity field and ring mass. {\it Science}, 364, Issue 6445, id. aat2965. \indent
 
Jacobson, R. A., (2009). The Orbits of the Neptunian Satellites and the Orientation of the Pole of Neptune. {\it AJ}, 137, 4322.  \indent
 
Jacobson, R.A. et al., (2006). The gravity field of the saturnian system from satellite observations and spacecraft tracking data. {\it ApJ}, 132, 2520.  \indent
 
Jacobson, R.A., (2014). The Orbits of the Uranian Satellites and Rings, the Gravity Field of the Uranian System, and the Orientation of the Pole of Uranus. {\it AJ}, 148, 76.  \indent
 
Karkoschka, E.\& Tomasko, M.~G., (2011). The haze and methane distributions on Neptune from HST-STIS spectroscopy. {\it Icarus}, 211, 780.  \indent
 
Kaspi, Y., Showman, A.~P., Hubbard, W.~B., Aharonson, O.~\& Helled, R., (2013). Atmospheric confinement of jet streams on Uranus and Neptune. {\it Nature}, 497, 344.  \indent

Kegerreis J. A.~et al., (2018). Consequences of Giant Impacts on Early Uranus for Rotation, Internal Structure, Debris, and Atmospheric Erosion. {\it ApJ}, 861, 52.  \indent
 
Kegerreis, J.~A., and 6 colleagues, (2019). Planetary giant impacts: convergence of high-resolution simulations using efficient spherical initial conditions and SWIFT. {\it MNRAS}, 487, 5029.  \indent
 
Kubo-Oka, T.~\&  Nakazawa, K., (1995). Gradual increase in the obliquity of Uranus due to tidal interaction with a hypothetical retrograde satellite.  {\it Icarus}, 114, 21. \indent 

Kurosaki, K.~\& Inutsuka, S., (2018). The Exchange of Mass and Angular Momentum in the Impact Event of Ice Giant Planets: Implications for the Origin of Uranus. {\it ApJ}, 157, 13 pp.   \indent
 
Kurosaki, K.~\& Inutsuka, S., (2019). The Exchange of Mass and Angular Momentum in the Impact Event of Ice Giant Planets: Implications for the Origin of Uranus. {\it AJ}, 157, 13.  \indent
%
  
Lindal, G. F., (1992). The Atmosphere of Neptune: an Analysis of Radio Occultation Data Acquired with Voyager 2. {\it AJ}, 103, 967.  \indent
  
Lindal, G.~F., Lyons, J.~R., Sweetnam, D.~N., Eshleman, V.~R., Hinson, D.~P., Tyler, G.~L., (1987). The atmosphere of Uranus: Results of radio occultation measurements with Voyager 2. {\it Journal of Geophysical Research}, 92, 14987.   \indent

Leconte, J., Selsis, F., Hersant, F., \& Guillot, T., (2017).  Condensation-inhibited convection in hydrogen-rich atmospheres: Stability against double-diffusive processes and thermal profiles for Jupiter, Saturn, Uranus, and Neptune, {\it Astronomy and Astrophysics}, 598, A98. \indent
 
Li, L., Jiang, X., West, R. A., Gierasch, et al (2018). Less absorbed solar energy and more internal heat for Jupiter. {\it Nature Communications}, 9, 3709.  \indent
 
Mankovich, C., Marley, M. S., Fortney, J.~J.~\& Movshovitz, N. (2019). Cassini Ring Seismology as a Probe of Saturn’s Interior. I. Rigid Rotation. {\it ApJ}, 871, 1.\\
 
Mankovich, C. (2020). Saturn's Rings as a Seismograph to Probe Saturn's Internal Structure. {\it AGU Advances}, 1, e00142.  \indent
 
Marley,  M.~S., G{\'o}mez P.~ \& Podolak, M.~(1995). Monte Carlo interior models for Uranus and Neptune. {\it GJR}, 100,  23349.  \indent

Masters, A., Achilleos, N., Agnor, C. B., et al., (2014), Neptune and Triton: Essential pieces of the Solar System puzzle {\it Planetary \& Space Science}, 104, 108.  \indent

McKinnon, W. B., Stern, S. A., Weaver et al., (2017). Origin of the Pluto-Charon system: Constraints from the New Horizons flyby.  {\it Icarus}, 287, 2. \indent
 
Millot, M., Coppari, F., Rygg, J.~R., Correa Barrios, A., Hamel, S., Swift, D. C.~\& Eggert, J.~H., (2019). Nanosecond X-ray diffraction of shock-compressed superionic water ice. {\it  Nature}, 569, 251. \indent
 
\clearpage 
Mousis, O.~et al., (2018). 
Scientific rationale for Uranus and Neptune in situ explorations. 
{\it Planetary and Space Science}, 155, 12.  \indent
 
Movshovitz, N., Fortney, J.~J., Mankovich, C., Thorngren, D.~\& Helled, R., (2020). Saturn's Probable Interior: An Exploration of Saturn's Potential Interior Density Structures. {\it ApJ}, 891, 109.  \indent 

Mousis, O., Aguichine, A., Helled, R., 
Irwin, P.~\& Lunine, J.~I. (2020). The role of ice lines in the
formation of Uranus and Neptune. 
{\it RSTA}, submitted.  \indent
 
Nesvorný, D., Vokrouhlický, D.~\&  Morbidelli, A., (2007). Capture of Irregular Satellites during Planetary Encounters. {\it AJ},  133, 1962.  \indent
 
Nettelmann, N., Fortney, J. J., Moore, K.~\& Mankovich, C.~(2015).  An exploration of double diffusive convection in Jupiter as a result of hydrogen-helium phase separation. 
{\it MNRAS}, 447, 3422.  \indent
 
Nettelmann, N., Helled, R., Fortney, J.J.~\& Redmer, R., (2013). 
\newblock {New indication for a dichotomy in the interior structure of Uranus
  and Neptune from the application of modified shape and rotation data}.
\newblock {\em Planet.~Sp.~Sci.}, 77, 143.  \indent
%
 
Nettelmann, N., Wang, K., Fortney, J.~J.~, Hamel, S., Yellamilli, S., Bethkenhagen, M.~\& Redmer, R., (2016). Uranus evolution models with simple thermal boundary layers. {\it Icarus}, {275}, 107.  \indent
 
{\"O}berg, K.~I., Murray-Clay, R.~\& Bergin, E.~A., (2011). The effects of snowlines on C/O in planetary atmospheres. {\it ApJ}, 743, L16.  \indent
 
Pearl, J. C., Conrath, B. J., Hanel, R. A., Pirraglia, J. A., \& Coustenis, A., (1990). {\it Icarus}, 84, 12.  \indent
 
Pearl, J.~C., Conrath, B.~J., (1991). The albedo, effective temperature, and energy balance of Neptune, as determined from Voyager data.\ Journal of Geophysical Research 96, 18921.  \indent
 
Podolak, M., Hubbard, W.~B.~\& Stevenson, D.~J. Model of Uranus interior and magnetic field. In:
Uranus, vol 2961. University of Arizona Press, Tucson, (1991).  \indent

Podolak, M., Weizman, A.~\& Marley, M., (1995). Comparative models of Uranus and Neptune. {\it PSS}, {43}, 1517.  \indent
%
 
Podolak, M.~\& Helled, R., (2012). What do we really know about Uranus and Neptune? {\it ApJ}, {759}, L7.  \indent
 
Podolak, M.~\, Helled, R.~\& Schubert, G., (2019). Effect of non-adiabatic thermal profiles on the inferred compositions of Uranus and Neptune. {\it MNRAS}, 487, 2653.  \indent
%
\clearpage
Redmer, R., Mattsson, T. R., Nettelmann, N., et al., (2011). The phase diagram of water and the magnetic fields of Uranus and Neptune, {\it Icarus}, 211, 798.  \indent 
 
Reinhardt, C., Chau, A., Stadel, J.~\& Helled, R., (2020). Bifurcation in the history of Uranus and Neptune: the role of giant impacts. {\it MNRAS},  492, 5336.  \indent
 
Rogoszinski, Z.~\& Hamilton, D.~P., (2020). Tilting Ice Giants with a Spin-Orbit Resonance. {\it ApJ}, 888, 60.  \indent
 
Rowe, J. F., Gaulme, P., Lissauer, J.~J.,
et al., (2017). Time-series Analysis of Broadband Photometry of Neptune from K2. {\it AJ}, 153, 149.  \indent
 
Safronov, V. S., (1966). Sizes of the Largest Bodies Falling onto the Planets during Their Formation, {\it Soviet Astronomy}, 9, 987.  \indent
 
Scheibe, L., Nettelmann, N.~\& Redmer. R., (2019). 
\newblock {Thermal evolution of Uranus and Neptune I: adiabatic models}.
\newblock {\em A\&A}, 632, A70.  \indent
 
Stevenson, D. J., (1986). The {{Uranus}}-{{Neptune Dichotomy}}: The {{Role}} of {{Giant Impacts}}, {\it Lunar and Planetary Science Conference}, 17, 1011--1012.  \indent
 
Stevenson, D. J., (2020). Jupiter's Interior as Revealed by Juno. {\it Annual Review of Earth and Planetary Sciences}, 48, 18.1. \indent

Soubiran, F., Militzer, B., Driver, K.~P., Zhang, S., (2017). Properties of hydrogen, helium, and silicon dioxide mixtures in giant planet interiors. {\it Physics of Plasmas}, 24, 041401.  \indent
 
Soubiran, F.~\& Militzer, B., (2015). Miscibility Calculations for Water and Hydrogen in Giant Planets.{\it ApJ}, 806, 228.  \indent

Sromovsky, L.~A., Fry, P.~M., Kim, J.~H., (2011). Methane on Uranus: The case for a compact CH $_{4}$ cloud layer at low latitudes and a severe CH $_{4}$ depletion at high-latitudes based on re-analysis of Voyager occultation measurements and STIS spectroscopy. {\it Icarus}, 215, 292.  \indent

Teanby, N. A. , Irwin, P. G. J., Moses, J. I.~\& Helled, R.,(2020).
Neptune: ice or rock giant? {\it RSTA}, submitted.  \indent

Thommes, E.~W., Duncan, M.~J., Levison, H.~F., (1999). The formation of Uranus and Neptune in the Jupiter-Saturn region of the Solar System.  {\it Nature}, 402, 635.  \indent

Tsiganis, K., Gomes, R., Morbidelli, A., Levison, H.~F., (2005). Origin of the orbital architecture of the giant planets of the Solar System.  {\it Nature}, 435, 459. \indent

Tyler, G. L., Sweetnam, D. N., Anderson, J. D., et al.,  (1989). 
Voyager radio science observations of Neptune and Triton. {\it Science}, 246, 1466.  \indent
%

Vazan, A., Kovetz, A., Podolak, M., Helled, R., (2013). The effect of composition on the evolution of giant and intermediate-mass planets.{\it MNRAS}, 434, 3283.  \indent

Vazan, A., Helled, R.~\& Guillot, T., (2018). Jupiter's evolution with primordial composition gradients. {\it A\&A}, 610, L14. \indent

Vazan, A.~\& Helled, R., (2020). Explaining the low luminosity of Uranus: a self-consistent thermal and structural evolution. {\it A\& A}, 633, id.A50, 10 pp. \indent

Wahl S. et al., (2017). Comparing Jupiter interior structure models to Juno gravity measurements and the role of a dilute core. {\it Geophysical Research Letters} 44, 4649.  \indent

Canup, R.~M.~\& Ward, W.~R., (2006). 
A common mass scaling for satellite systems of gaseous planets. 
{\it Nature}, 441, 834.  \indent

Warwick, J.~W., and 21 colleagues, (1989). Voyager Planetary Radio Astronomy at Neptune. {\it Science}, 246, 1498.  \indent

Wong, M. H., Mahaffy, P. R., Atreya, S. K., Niemann, H. B.~\& Owen, T. C., (2004). Updated Galileo probe mass spectrometer measurements of carbon, oxygen, nitrogen, and sulfur on Jupiter. {\it Icarus}, 171, 153.
}

\clearpage

\begin{figure}
\centering
   \includegraphics[angle=0,height=7.5cm]{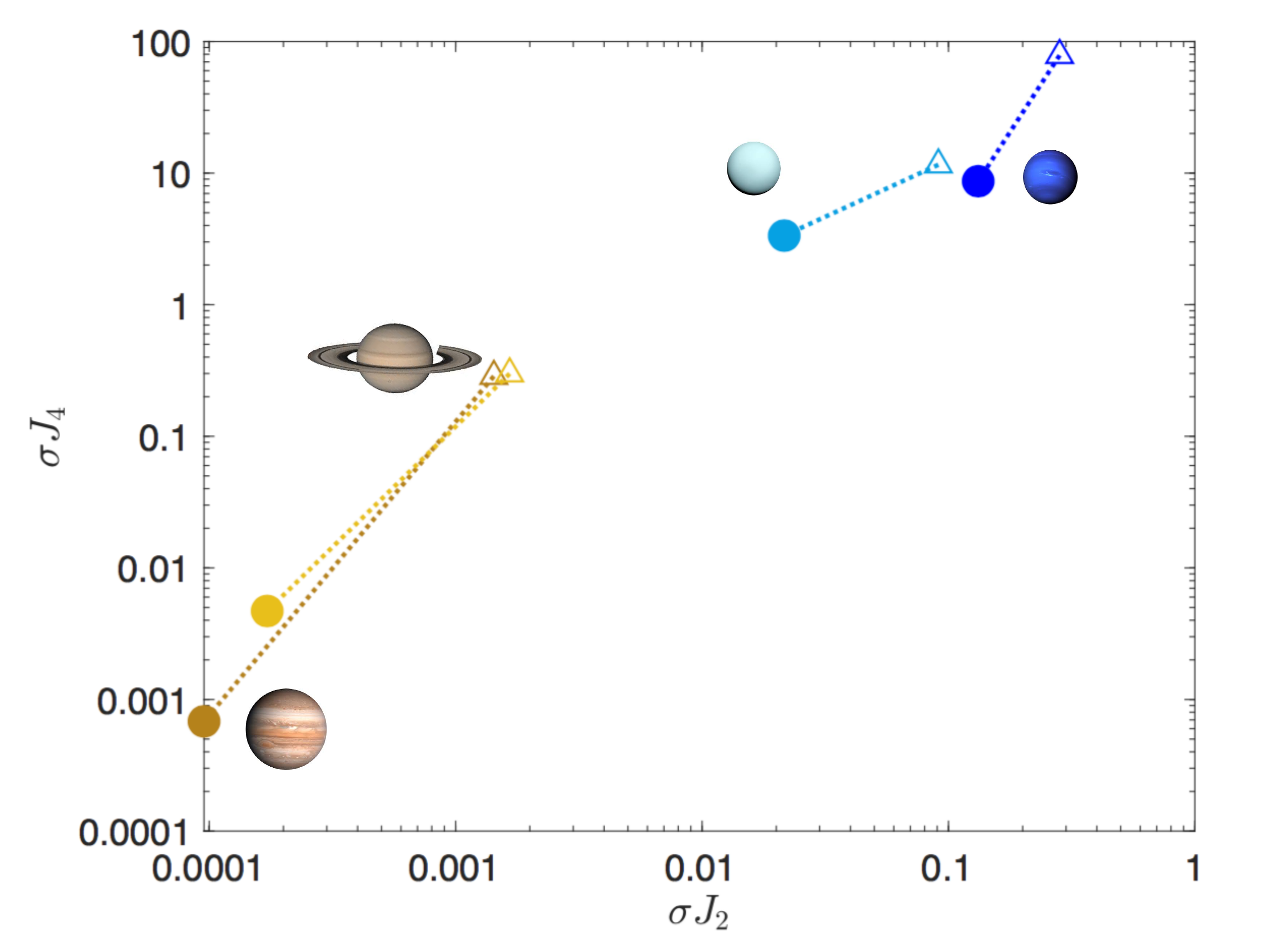}
\caption{
\footnotesize{
{\bf Gravity measurement uncertainties of the outer planets.} 
Shown are relative observational uncertainties in $J_2$ and $J_4$ of the four giant planets. The triangles and circles represent "old" and "new" data, respectively. 
"Old" data correspond to the latest data before
the most recent, i.e., "new" data.  The used numbers are summarized in Table 2. 
}
}
\end{figure}

\begin{figure}[h]
\centering
   \includegraphics[angle=0,height=7.15cm]{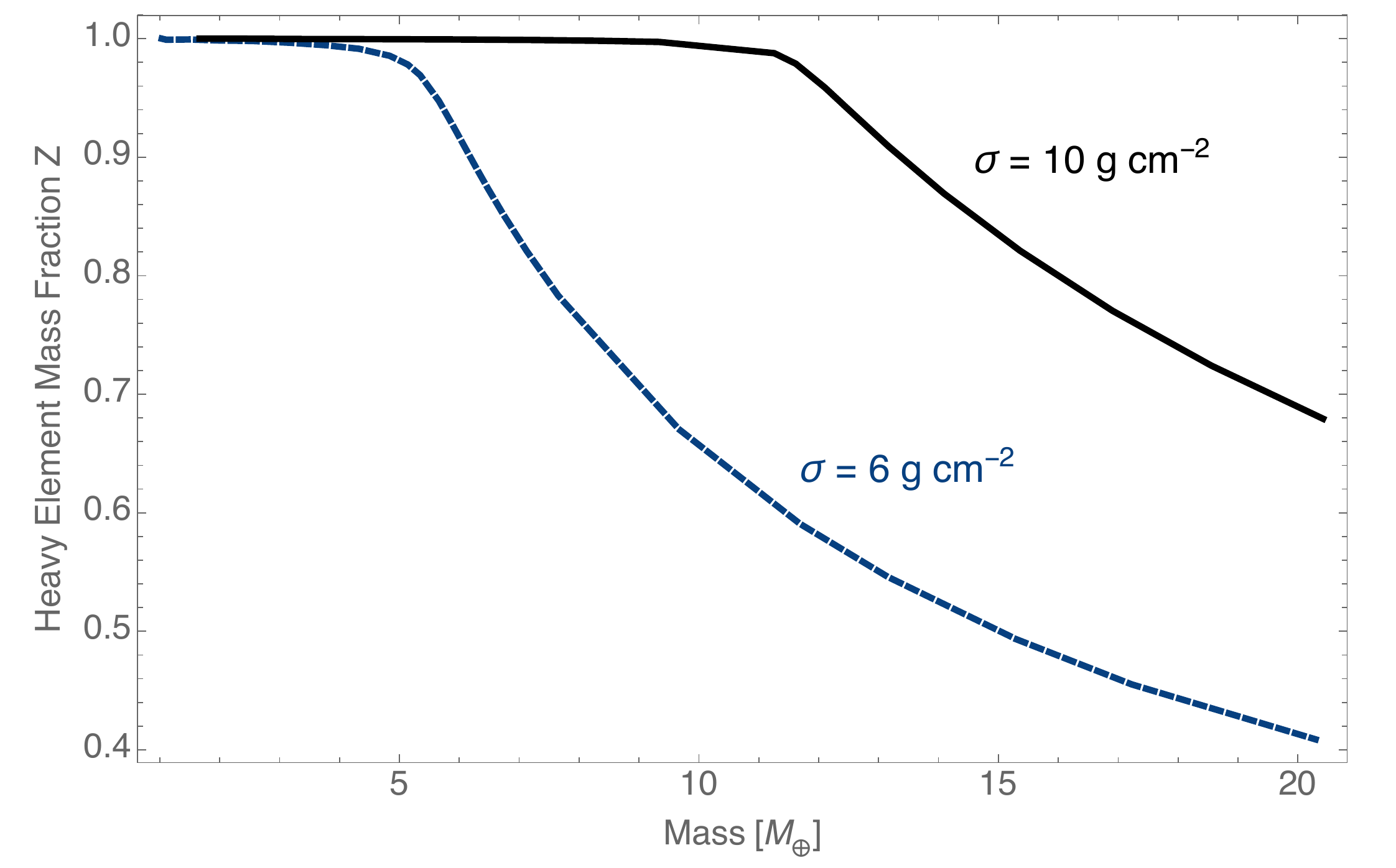}
\caption{
\footnotesize{
{\bf Heavy element mass fraction $Z$ vs.~planetary mass up to a mass of 20 M$_{\oplus}$.} 
The solid-black and dotted-blue curves correspond to formation models assuming  solid surface densities of 10 g cm$^{-2}$ and 6 g cm$^{-2}$, respectively. This demonstrates the dependence of the planetary composition and primordial internal structure on the relative accretion rates (see Helled \& Stevenson 2017 for details). 
}}
\label{fig:1}       
\end{figure}

\newpage
\begin{figure}
   \includegraphics[angle=0,height=6.5cm]{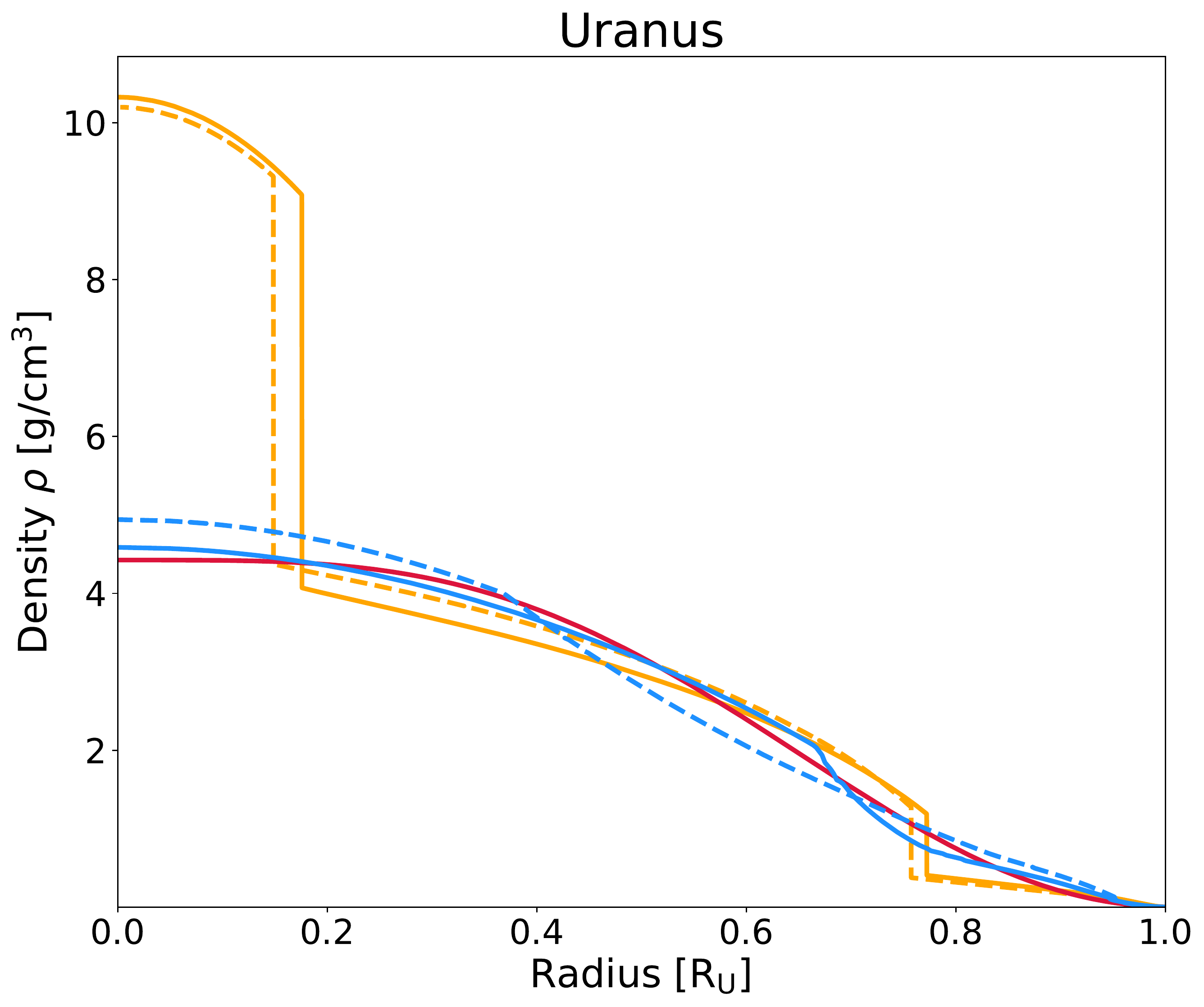}
  \includegraphics[angle=0,height=6.5cm]{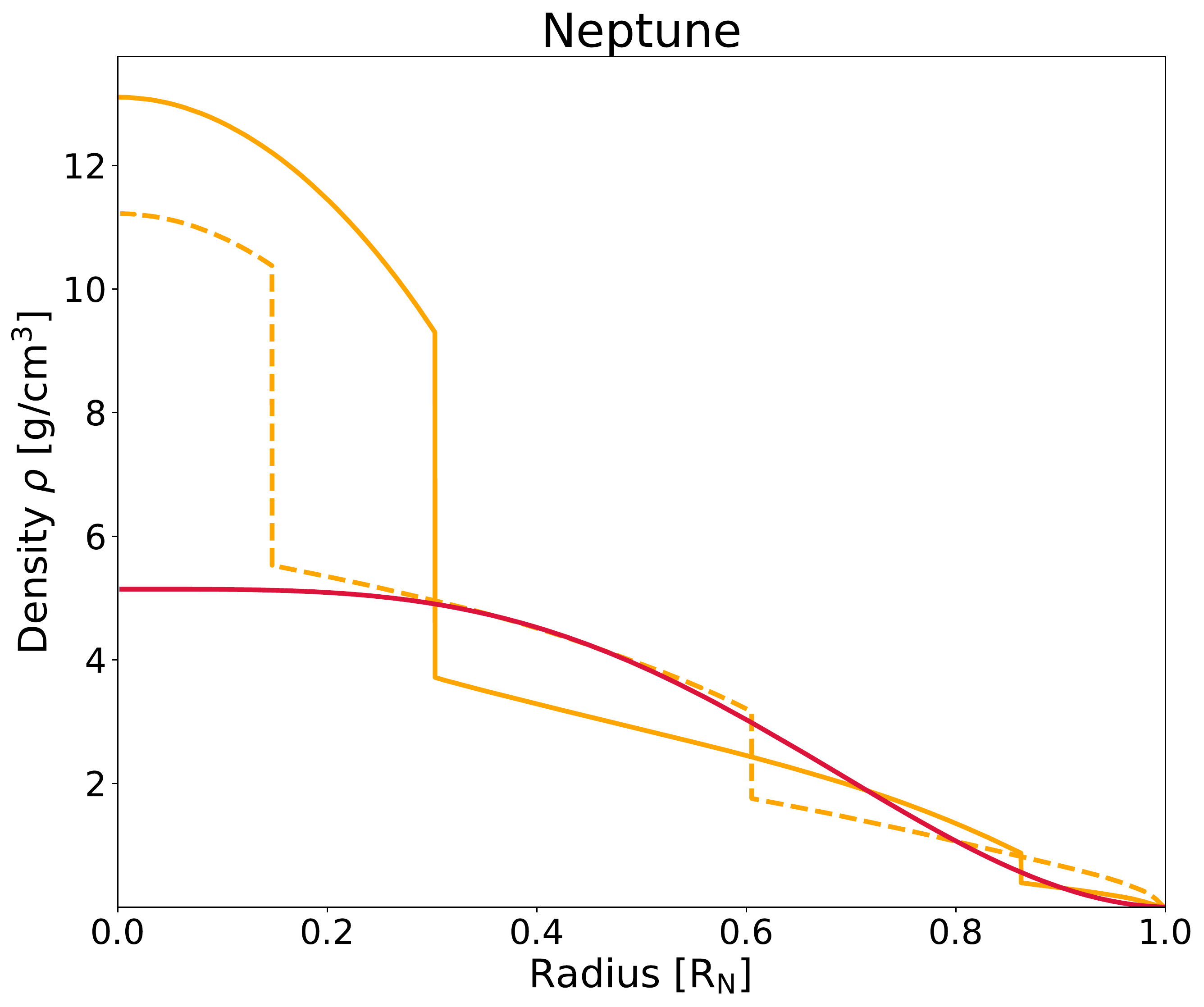}
\caption{
\footnotesize{{\bf Density profiles of Uranus and Neptune.} The colors correspond to different studies. For Uranus,  the solid and dashed orange curves are models U1 and U2 from Nettelmann et al.~(2013), the solid red is from  Helled et al.~(2011), and the solid and dashed blue lines are models V2 and V3 from Vazan \& Helled, 2020, respectively. For Neptune, the solid and dashed orange lines are models N1 and N2b from Nettelmann et al.~(2013), and the solid red is from  Helled et al.~(2011).}}
\label{fig:1}       
\end{figure}

\begin{figure*}
   \includegraphics[angle=0,height=6cm]{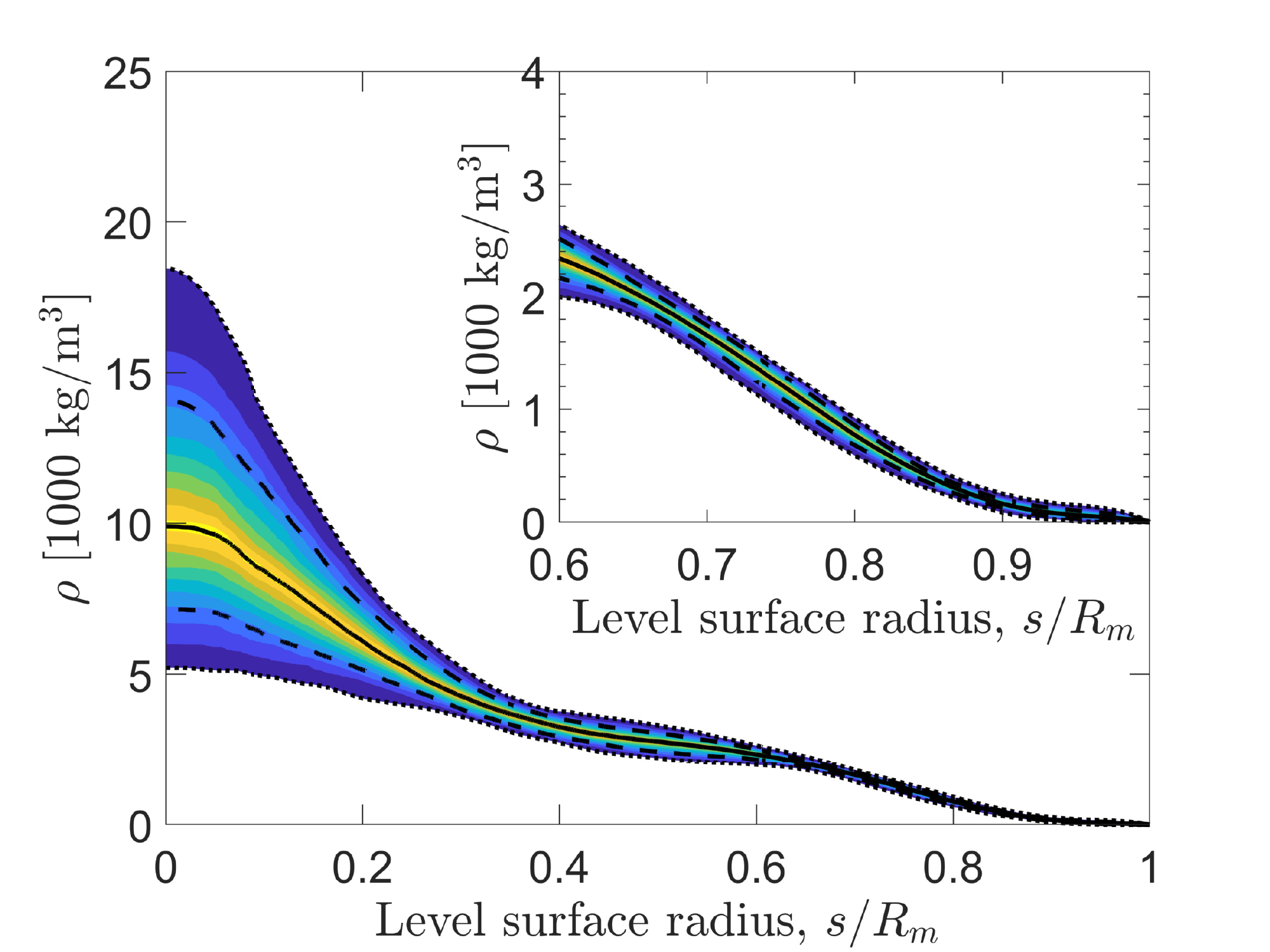}
  \includegraphics[angle=0,height=6cm]{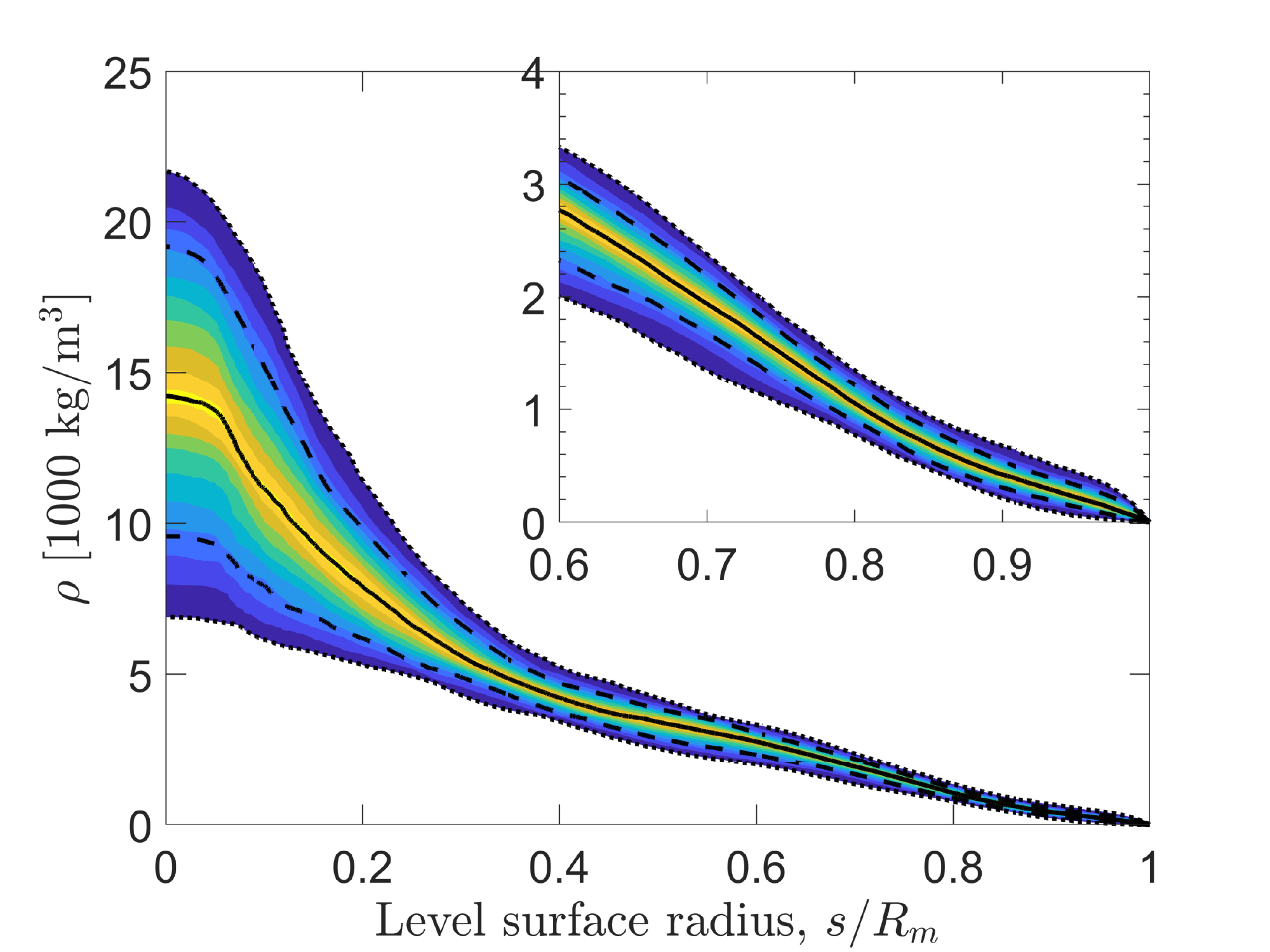}
\caption{
\footnotesize{{\bf Empirical density profiles of Uranus (left) and Neptune (right) derived from 8th-order polynomials.} Visualization of the posterior probability distribution of Uranus and Neptune interior density profiles.  The thick black line is the sample-median of density on each level surface. The dashed lines mark the the 16th and 84th percentiles and the dotted
lines mark the 2nd and 98th percentiles; between the lines percentile value is  color-coded.  Inset shows a zoom-in in outer layers.  Uranus is better-constrained than Neptune}.}
\label{fig:naor}       
\end{figure*}

\begin{figure}
\centering
   \includegraphics[angle=0,height=6.5cm]{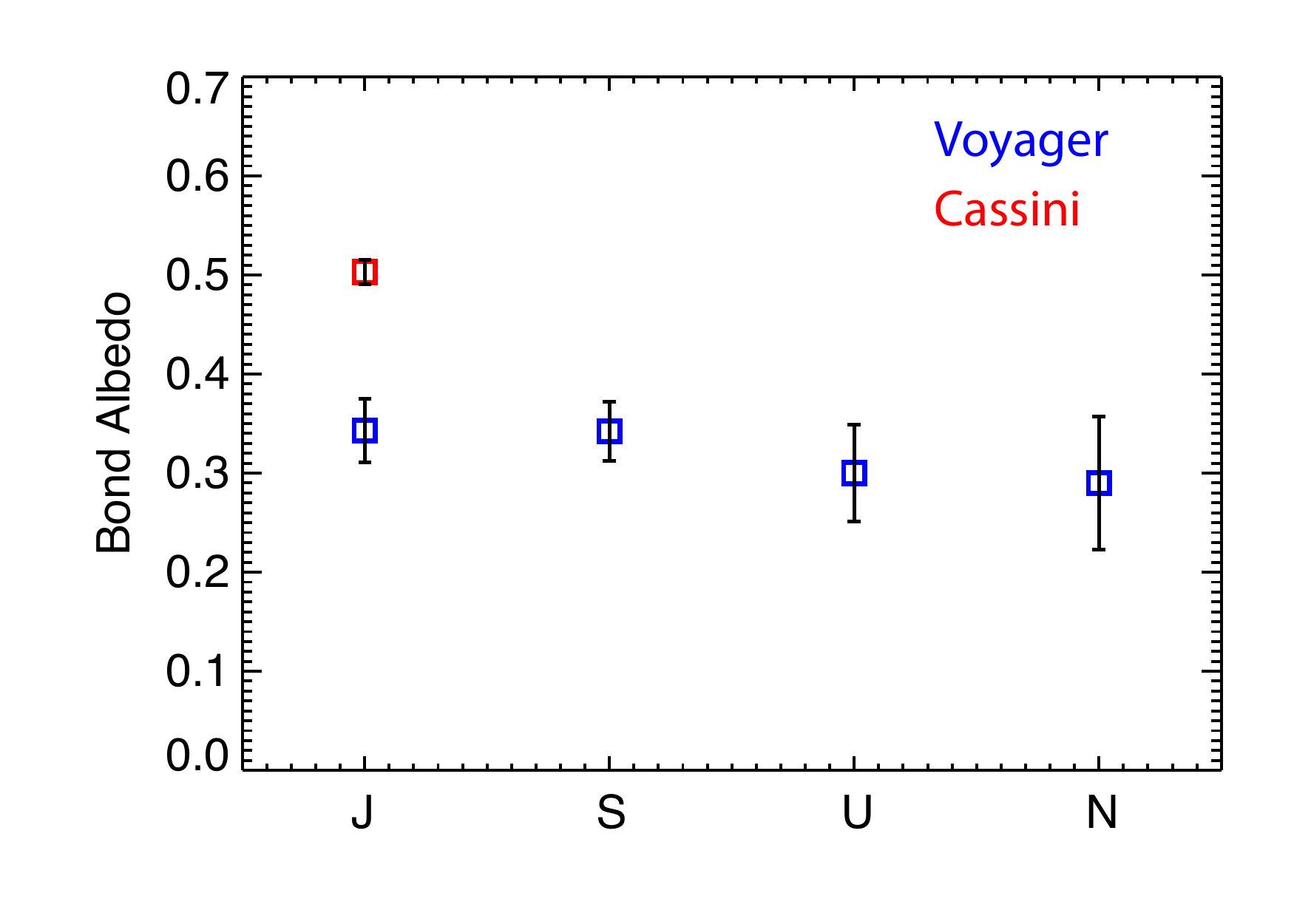}
\caption{
\footnotesize{{\bf Bond albedo determinations of the giant planets.} 
Shown are the Bond albedos of Jupiter, Saturn, Uranus, and Neptune, from left to right.  Values from Voyager data are shown in blue with $1 \sigma$ error bars.  The data for Jupiter, Saturn, Uranus and Neptune are based on Hanel et al., (1981), Hanel et al., (1983), Peal et al.~(1990), and Pearl \& Conrath, (1991), respectively. 
The red value for Jupiter is from \emph{Cassini} fly-by data (Li et al., 2018). 
}}
\label{fig:alb}       
\end{figure}

\end{document}